\documentclass[12pt]{iopart}

\def\const{{\rm const}}

\begin{document}

\title{Different faces of the phantom}

\author{K.A. Bronnikov\footnote{E-mail: kb20@yandex.ru. Permanent address:
	Center for Gravitation and Fundamental Metrology, VNIIMS, 46
	Ozyornaya St., Moscow, Russia; Institute of Gravitation and
	Cosmology, PFUR, 6 Miklukho-Maklaya St., Moscow 117198, Russia.},
     J.C. Fabris\footnote{E-mail: fabris@cce.ufes.br} and
     S.V.B. Gon\c{c}alves\footnote{E-mail: sergio@cce.ufes.br}}

\address{Departamento de F\'{\i}sica, Universidade Federal do Esp\'{\i}rito
	Santo, Vit\'oria, ES, Brazil}

\begin{abstract}
The SNe type Ia data admit that the Universe today may be dominated by some exotic matter with negative pressure
violating all energy conditions. Such exotic matter is called {\it phantom
matter} due to the anomalies connected with violation of the energy
conditions. If a phantom matter dominates the matter content of the
universe, it can develop a singularity in a finite future proper time. Here
we show that, under certain conditions, the evolution of perturbations of
this matter may lead to avoidance of this future singularity (the Big Rip).
At the same time, we show that local concentrations of a phantom
field may form, among other regular configurations, black holes with
asymptotically flat static regions, separated by an event horizon from an
expanding, singularity-free, asymptotically de Sitter universe.

\end{abstract}


\section{Introduction}

The evidence for an accelerating expanding phase of the universe seems to be
robust \cite{SN}. If this is the case, the deceleration parameter $q = -
\ddot a\,a/\dot a^2$, must be negative, implying that the matter dominating
the matter content of the universe, if described in terms of a perfect
fluid, must have an equation of state $p = w\rho$, with $w = - 1/3$. Hence,
the strong energy condition must be violated. More recently, there has been
claims that the observational data favour an equation of state with $w < -
1$ \cite{PHANTOM}. Matter with such an equation of state violates all the
energy conditions. This kind of matter can be represented, in a more
fundamental way, by a self-interacting scalar field, whose kinetic energy
appears with the "wrong sign". For this reason, it is more popularly called
a {\it phantom field}. In the usual hydrodynamics representation, a matter
with this equation of state is unstable. However, such an
instability may disappear when a fundamental description is employed, for
example, using a self-interacting scalar field, as stated before.

If phantom matter dominates the matter content of the Universe, a future
singularity may develop in a finite proper time since its density grows as
the universe expands. Such a singularity inevitably appears in an
isotrpic Universe if $w=\const < -1$). This possible future singularity
has been named a {\it Big Rip}, leading to the notion of a {\it phantom
menace} \cite{caldwell}.  However, it must be stressed that the possibility
that phantom matter dominates the universe today is yet a matter of debate:
the observational data, mainly those coming from the supernovae type Ia of
high redshift, lead to different conclusions depending on how the sample is
selected, and even how the statistical analysis is performed
\cite{padmana1}. In any case, the possibility that a phantom matter has
something to do with the actual universe must be taken seriously.

The goal of the present work is twofold. First, we intend to analyse the
evolution of scalar perturbations for phantom matter. This is an important
point, since we are dealing with matter with negative pressure, and
instabilities may develop, mainly at small scales \cite{jerome}.  It is
possible to get rid of these instabilities if a scalar field representation
is used: the behaviour of the perturbations on small scales are quite
sensitive to the description used for the matter \cite{nazira}. On the other
hand, the behaviour of the perturbations at large scales are quite
insensitive to the description employed. We will show, using a
self-interacting scalar field {\rm simulating a perfect fluid with $w
=\const$}, that phantom matter may lead to growing perturbations at large
scales if the pressure is negative enough. This may lead to a very
inhomogenous universe deep in the phantom era, and such inhomogeneties may
lead to avoidance of a Big Rip.

Besides this perturbative analysis of a cosmological scenario where the
phantom matter dominates the matter content of the universe, we will study
local configurations with spherical symmetry. In this case, the results are
still more unexpected. The fact that the kinetic term has a "wrong" sign,
may lead to a minimum of the radius of coordinate 2-spheres, so that a central
singularity is avoided by having no center at all, as is the case with
wormholes. However, such configurations may contain one or two Killing
horizons, and, among others, it is possible to have configurations where
there is a static, asymptotically flat region which is separated by an
event horizon from an expanding singularity-free, Kantowski-Sachs type
universe. It is thus a black hole in which an explorer may survive after
crossing the horizon. 

\section{Evolution of scalar perturbations}

When a barotropic fluid with the equation of state $p = w\rho$ is
introduced in the Einstein's equations, with a Friedmann-Robertson-Walker
flat line element $ds^2 = a^2(d\eta^2 - dx^2 - dy^2 - dz^2)$, $\eta$ being
the conformal time, the conservation equation
\begin{equation}
\dot\rho + 3\frac{\dot a}{a}(\rho + p) = 0 \quad ,
\end{equation}
leads, in case $w = \const$, to $\rho \propto a^{-3(1 + w)}$.
Inserting this into the Friedmann equation
\begin{equation}
\biggr(\frac{a'}{a}\biggl) = \frac{8\pi G}{3}\rho\,a^2 \quad ,
\end{equation}
we find that the scale factor behaves as $a \propto \eta^\frac{2}{1 + 3w}$.
One important feature of this solution that must be stressed in order to
understand the behaviour at perturbative level, is the
character of the "future". If  $w > - 1/3$, when all energy conditions
hold, $\eta \rightarrow \infty$ means $a
\rightarrow \infty$; when $w < - 1/3$, on the other hand, the universe is
expanding as $\eta \rightarrow 0_-$.

In general, fluids with negative pressure contain, at a perturbative level,
decreasing modes at large scales and unstable models at small scales
\cite{jerome}. The instabilities at small scales must not be taken so
seriously, since it must be due mainly to the hydrodynamical approximation
\cite{nazira}. It is possible to use a more fundamental representation for
such exotic matter by considering a self-interacting scalar field, which
reproduces, from the point of view of the background behaviour, the
hydrodynamical approach employed until now. A scale factor which evolves as
$a \propto \eta^\frac{2}{1 + 3w}$, with $w < - 1$, can be achieved by
considering a self-interacting minimally coupled scalar field, such that,
\begin{equation}
\label{potential}
V(\phi) = V_0\exp\biggr(\pm\sqrt{-3(1 + w)}\phi\biggl)  \quad ,
\quad \phi = \pm 2\frac{\sqrt{-3(1 + w)}}{1 + 3w}\ln\eta \quad .
\end{equation}
A similar model can be constructed when $w > - 1$ by just changing the sign
of the term inside the square roots.

For gravity minimally coupled to a (self-interacting) scalar field,
the equations for the perturbed quantities reduce to a single equation for the metric perturbed
function $\Phi$, called Bardeen's potential, which is \cite{brand}
\begin{equation}
\Phi'' + 2\biggr\{H - \frac{\phi''}{\phi'}\biggl\}\Phi'
+ \biggr\{k^2 + 2\biggr[H'- H\frac{\phi''}{\phi'}\biggl]\biggl\}
\Phi = 0 \quad .
\end{equation}
Using the background expressions for $H$ and $\phi$, this equation becomes,
\begin{equation}
\Phi'' + 2 \frac{3(1 + w)}{1 + 3w}\frac{\Phi'}{\eta} + k^2\Phi = 0 \quad ,
\end{equation}
with the solutions
\begin{equation}
\Phi = (k\eta)^{-\nu}\biggr\{c_1(k)J_\nu(k\eta)
+ c_2(k)J_{-\nu}(k\eta)\biggl\} \quad , \quad
\forall\, w \quad ,
\end{equation}
where $\nu = (5 + 3w)/[2(1 + 3w)]$ and $k$ is the wavenumber of the perturbations, resulting from a plane wave expansion of the
spatial part of the perturbed quantities.

In the small-scale asymptotic limit defined by $k\eta \ll  1$, the solutions
behave as in the hydrodynamical representation:
\begin{equation}
\Phi \propto c_1 + c_2(k\eta)^{-2\nu} \quad .
\end{equation}
In all cases, there is a constant mode. However, if $\omega > - 5/3$ the
second mode decreases as the universe expand; but there is a growing mode
when $w \leq - 5/3$, what can lead to formation of large inhomogeneties.
There is an asymptotic logarithmic divergence for $w = - 5/3$.  Using, on
the other hand, the asymptotic expression for the Bessel functions for large
values of the argument $k\eta \gg  1$, the potential can be expressed as
\begin{equation}
\Phi \sim (k\eta)^{-\frac{1 + w}{1 + 3w}}\cos(k\eta + \delta) \quad ,
\end{equation}
$\delta$ being a phase. It is easy to verify that for $w > - 1$, the
potential oscillates with decreasing amplitude, while for $w < - 1$, the
potential oscillates with increasing amplitude. Hence, for $w < - 5/3$ the
phantom field may exhibit instability at large and small scales. However, it
must be stressed that the behaviour at small scale is quite model-dependent,
and another field representation of the phantom field can modify the
conclusions at small scales, like considering the phantom field as a ghost
condensation \cite{piazza} or a tachyon \cite{padmana2}. But, at large
scales, it seems that there is always a growing mode, for $w \leq - 5/3$,
irrespective of the representation chosen.

It is fundamental now to understand the meaning of small and large
asymptotic limits, $k\eta \ll  1$ or $k\eta \gg  1$ respectively. We normalize
the scale factor by fixing $a_0 = 1$ at the present time. This implies that
the Hubble parameter, expressed in terms of the cosmic time, is given by
${\cal H}_0 = \frac{\dot a}{a}|_{t=t_0} = \frac{a'}{a^2}|_{\eta=\eta_0} =
\frac{2}{|1 + 3w|}\frac{1}{|\eta_0|}$, where $\eta_0$ is the conformal time
today. Hence, $\eta_0 \sim {\cal H}_0^{-1} = l_H$ (with $c = 1$). This
implies that the separation point between the large and
small scale regimes is given by the Hubble length $l_H$. When $w > - 1/3$,
the conformal time increases as the Universe expands, implying that, as time
goes on, more and more modes satisfy the condition $k\eta \gg 1$, which can
be re-expressed by saying that the modes enter the Hubble horizon as time
goes on.  The opposite occurs when $w < - 1/3$, and as time goes on, more
and more physical modes satisfy the condition $k\eta \ll  1$, the usual
situation of an accelerated expansion phase. In the interval $- 5/3 < w < -
1/3$, these modes that are stretched outside the Hubble horizon are frozen
or decay, and there is no danger for homogeneity.  But, for $w < - 5/3$,
these modes begin to become strongly unstable and homogeneity can be
destroyed.  The above results can be re-expressed using, as a reference
parameter, the Hubble horizon as a function of time.  In fact, using the
expression for the Hubble length at any time, we have $l_H(\eta) = \frac{|1
+ 3w|}{2}|\eta|^\frac{3(1 + w)}{1 + 3w}|\eta_0|^\frac{-2}{1 + 3w}$ and the
argument of the Bessel functions written above can be expressed as $k\eta
\sim k\,[l_H(\eta)]^\frac{1 + 3w}{3(1 + w)}$. For the phantom field,
$l_H(\eta)$ decreases as the Universe expands. Hence, for $w < - 5/3$ more
and more modes go out of the Hubble horizon and begins to grow, enchancing
the inhomogeneity. Notice that for $w = - 1$ (cosmological constant case),
the Hubble horizon remains constant.

Even if the calculations exposed here were done for the flat case, they also
apply to the open and closed cases \cite{deborah}. These results have
been obtained with the potential (\ref{potential}) that reproduces the equation of state
$w = \const$ in an isotropic universe. It should be mentioned that other
froms of the potential may not lead to a Big Rip even in the isotropic case:
e.g., if $V$ is bounded above, a phantom-dominated universe evolves, in
general, toward a de Sitter attractor solution \cite{fara05}.

\section{Local configurations}

Let us consider now the Hilbert-Einstein Lagrangian coupled to a
self-interacting scalar field with an unspecified sign for the kinetic term.
For the moment, we do not specify the potential. Considering the
spherically symmetric metric written in the form
\begin{equation}
ds^2 = Adt^2 - A^{-1}d\rho^2 - r(\rho)^2d\Omega^2 \quad,
\end{equation}
$\rho$ being the radial variable, we find the following set of coupled
differential equations:
\begin{eqnarray}                                                         \label{ne1}
    (A\,r^2\,\phi')' =  \epsilon r^2\,V_\phi \quad ,
\\                                                            \label{ne2}
    (A'\,r^2)' =  - 2V\,r^2 \quad ,
\\                                                            \label{ne3}
    2r''/r =  - \epsilon\,\phi'^2\quad  ,
\\                                                            \label{ne4}
    (r^2)''\,A - A''\,r^2 =  2\quad .
\end{eqnarray}
If $\epsilon = 1$ we have a "normal" scalar field, while if $\epsilon = - 1$ we have
a phantom scalar field.
 Equation (\ref{ne4}) is once integrated giving
\begin{equation}                                                          \label{ne4i}
    (A/r^2)' = 2 (\rho_0 - \rho)/r^4 \quad , \quad \rho_0 = \mbox{const} \quad .
\end{equation}
We will summarize the possible configurations in what follows without
solving explicitly the equations (\ref{ne1}-\ref{ne4}) \cite{kirill}.

Let us indicate the possible kinds of nonsingular solutions without
    restricting the shape of $V(\phi)$. Assuming no pathology at
    intermediate $\rho$, regularity is determined by the system behavior at
    the ends of the $\rho$ range. The latter may be classified as a regular
    infinity ($r \rightarrow \infty$), which may be flat, de Sitter or AdS, a regular center $r \rightarrow 0$, and the intermediate case $r \rightarrow r_0 > 0$.
Suppose we have a regular infinity as $\rho \rightarrow \infty$, so that
    $V \rightarrow V_+ = \mbox{const}$ while the metric becomes Minkowski (M), de Sitter
    (dS) or AdS according to the sign of $V_+$. In all cases $r \approx
    \rho$ at large $\rho$.

    For $\epsilon = +1$, due to $r'' \leq 0$, $r$ necessarily vanishes at some
    $\rho=\rho_c$, which means a center, and the only possible regular
    solutions interpolate between a regular center and an AdS, flat or dS
    asymptotic; in the latter case the causal structure coincides with that
    of de Sitter space-time.

    For $\epsilon = -1$, there are similar solutions with a regular center,
    but due to $r''\geq 0$ one may obtain either
    $r \rightarrow r_0 = \mbox{const} > 0$ or $r \rightarrow \infty$ as $\rho \rightarrow -\infty$. In other
    words, all kinds of regular behavior are possible at the other end.
    In particular, if $r \rightarrow r_0$, we get $A \approx -\rho^2/r_0^2$, i.e., a
    cosmological region comprising a highly anisotropic Kantowski-Sachs cosmology (KS) with one scale
    factor ($r$) tending to a constant while the other ($A$) inflates. The
    scalar field tends to a constant, while $V(\phi) \rightarrow 1/r_0^2$.

    Thus there are three kinds of regular asymptotics at one end,
    $\rho\rightarrow \infty$ (M, dS, AdS), and four at the other, $\rho \rightarrow  -\infty$:
    the same three plus $r \rightarrow r_0$, simply $r_0$ for short. (The asymmetry
    has appeared since we did not allow $r \rightarrow \mbox{const}$ as $\rho \rightarrow \infty$.
    The inequality $r''> 0$ forbids nontrivial solutions with two such
    $r_0$-asymptotics.) This makes nine combinations shown in Table 1.
    Moreover, each of the two cases labelled KS* actually comprises
    three types of solutions according to the properties of $A(\rho)$: there
    can be two simple horizons, one double horizon or no horizons between
    two dS asymptotics. Recalling 3 kinds of solutions with a regular
    center, we obtain as many as 16 qualitatively different classes of
    globally regular configurations of phantom scalar fields.

\begin{table}[h]
    \caption{Regular solutions with for $\epsilon = -1$. Each row
    corresponds to a certain asymptotic behavior as $\rho\rightarrow +\infty$, each
    column --- to $\rho\rightarrow -\infty$. The mark ``sym'' refers to combinations
    obtained from others by symmetry $\rho \leftrightarrow -\rho$.}
\begin{center}
\begin{tabular}{|c|c|c|c|c|}
\hline
	      &	  AdS    &    M     &   dS   &  $r_0$    \\
\hline
       AdS    &  wormhole     &   wormhole    &   black hole  &  black hole    \\
\hline
       M      &  sym     &   wormhole    &   black hole  &  black hole      \\
\hline
       dS     &  sym     &   sym    &   KS*  &   KS*     \\
\hline
\end{tabular}
\end{center}
\end{table}

\section{Conclusions}

Phantom matter implies violation of all energy conditions. In a
hydrodynamical representation, phantom matter is described by $p = w\rho$
with $w < - 1$. Phantom matter can be
described by a self-interacting scalar field with a "wrong" sign in the
kinetic term.  There is some evidence that the exotic matter responsable
for the actual phase of accelerated expansion of the universe may be a kind
of phantom field. In this work, we have studied some properties of a phantom
field, specifically with respect to the evolution of scalar perturbations
and with respect to local configurations.

We found that, under certain conditions, a universe dominated by a phantom
matter may develop high inhomogeneities even at large scales. Hence, after a
certain stage of its evolution, the hypothesis of homogeneity and isotropy
becomes no more valid.  As the big rip scenario depends on these hypothesis,
it is possible that a phantom universe brings in itself a mechanism of
avoiding a future singularity even in the case when $w = \const$ in
homogeneos and isotropic space-time.

In what concerns local configurations,
we find a wealth of nonsingular models among which of particular
interest are asymptotically flat black holes with an expanding universe
beyond the event horizon. This provides an interesting singularity-free
cosmological scenario: one may speculate that our Universe could appear from
collapse to such a phantom black hole in another, ``mother'' universe and undergo
isotropization (e.g., due to particle creation) soon after crossing the
horizon.

There are no similar configurations with a "normal"
scalar field.  In any case, violation of all energy condition
inevitably leads to completely new configurations.

{\bf Acknowledgments.}
  This work was supported by CNPq (Brazil); KB was also supported by
  DFG Project 436/RUS 113/807/0-1(R).

\end{document}